\documentclass[aps,prl,preprint]{revtex4-1}
\usepackage{graphicx}
\usepackage{epstopdf}
\usepackage{amssymb, amsmath}
\usepackage[usenames]{color}
\usepackage{bm}
\usepackage{multirow}
\usepackage{dcolumn}
\usepackage{hyperref}
\hypersetup{
	colorlinks=true,
	linkcolor=blue,
	filecolor=gray,      
	urlcolor=blue,
	citecolor=blue,
}

\begin{document}

\title{High-Temperature Quantum Anomalous Hall Insulators in Lithium-Decorated Iron-Based Superconductor Materials}

\author{Yang \surname{Li}$^{1,2,3}$}
\author{Jiaheng \surname{Li}$^{1,2,3}$}
\author{Yang \surname{Li}$^{1,2,3}$}
\author{Meng \surname{Ye}$^{1,2,3}$}
\author{Fawei \surname{Zheng}$^{4}$}
\author{Zetao \surname{Zhang}$^{1,2,3}$}
\author{Jingheng \surname{Fu}$^{1,2,3}$}
\author{Wenhui \surname{Duan}$^{1,2,3,5,6}$}
\email{duanw@tsinghua.edu.cn}
\author{Yong \surname{Xu}$^{1,2,3,7}$}
\email{yongxu@mail.tsinghua.edu.cn}

\affiliation{$^{1}$State Key Laboratory of Low Dimensional Quantum Physics and Department of Physics, Tsinghua University, Beijing, 100084, China \\
	$^{2}$Frontier Science Center for Quantum Information, Beijing, China \\
	$^{3}$Collaborative Innovation Center of Quantum Matter, Beijing 100084, China \\
	$^{4}$Laboratory of Computational Physics, Institute of Applied Physics and Computational Mathematics, Beijing 100088, China \\
	$^{5}$Institute for Advanced Study, Tsinghua University, Beijing 100084, China \\
	$^{6}$Beijing Academy of Quantum Information Sciences, Beijing 100193, China \\
	$^{7}$RIKEN Center for Emergent Matter Science (CEMS), Wako, Saitama 351-0198, Japan}

\begin{abstract}
Quantum anomalous Hall (QAH) insulator is the key material to study emergent topological quantum effects, but its ultralow working temperature limits experiments. Here, by first-principles calculations, we find a family of stable two-dimensional (2D) structures generated by lithium decoration of layered iron-based superconductor materials Fe$X$($X$ = S, Se, Te), and predict room-temperature ferromagnetic semiconductors together with large-gap high-Chern-number QAH insulators in the 2D materials. The extremely robust ferromagnetic order is induced by the electron injection from Li to Fe and stabilized by strong ferromagnetic kinetic exchange in the 2D Fe layer. While in the absence of spin-orbit coupling (SOC), the ferromagnetism polarizes the system into a half Dirac semimetal state protected by mirror symmetry, the SOC effect results in a spontaneous breaking of mirror symmetry and introduces a Dirac mass term, which creates QAH states with sizable gaps (several tens of meV) and multiple chiral edge modes. We also find a 3D QAH insulator phase featured by macroscopic number of chiral conduction channels in bulk LiOH-LiFe$X$. The findings open new opportunities to realize novel QAH physics and applications at high temperatures.
\end{abstract}

\maketitle

Seeking high-temperature quantum anomalous Hall (QAH) insulators is a compelling problem of condensed matter physics and material science~\cite{haldane2017nobel,hasan2010colloquium,qixialiang2011topological}. The QAH insulator, also called Chern insulator, is an exotic topological quantum state of matter that distinguishes from a normal insulator by a nonzero quantized Berry phase in the momentum space and can display quantized Hall conductance under zero magnetic field~\cite{haldane1988model}. Significantly distinct from the time-reversal invariant topological insulator (TI)~\cite{hasan2010colloquium,qixialiang2011topological}, the QAH state does not require any symmetry protection, and the time reversal symmetry is broken by spontaneous magnetization.
Such kind of magnetic topological quantum states provide a fertile playground to explore emergent quantum physics (e.g., topological magnetoelectric effects and Majorana fermions) and promising applications (e.g., low-power electronics and quantum computation)~\cite{haldane2017nobel,hasan2010colloquium,qixialiang2011topological}. However, as the QAH effect was only observable at liquid helium temperatures~\cite{changcuizu2013experimental,checkelsky2014trajectory,kou2014scale,changcuizu2015high,mogi2015magnetic}, its experimental observations and practical applications are seriously restricted.

The QAH insulator belongs to a specific type of two-dimensional (2D) ferromagnetic (FM) semiconductor,  which is featured by a topological energy gap opened by the spin-orbit coupling (SOC). Additionaly, in order to achieve high-temperature QAH state, both the FM coupling and SOC strength must be appreciable. However, high-temperature FM semiconductors are rare in nature~\cite{ando2006seeking} and for the 2D intrinsic FM semiconductors fabricated recently~\cite{gong2019two}, none of them features a nontrivial topology (except for MnBi$_2$Te$_4$~\cite{gongyan2019experimental,lijiaheng2019intrinsic,zhang2019topological}). Fundamentally, the search for high-temperature QAH insulators is challenged by the conflicting  material requirements of ferromagnetism and TI: The former prefers metallic systems composed of light 3$d$ elements, while the latter favors heavy elements for achieving strong SOC effects. The problem was partially circumvented by extrinsically introducing magnetic dopants to TI, or constructing magnetic-topological heterostructures~\cite{onoda2003quantized,liuchaoxing2008quantum,yurui2010quantized,changcuizu2013experimental,checkelsky2014trajectory,kou2014scale,changcuizu2015high,mogi2015magnetic}, or intercalating magnetic layers into heavy-element materials to realize intrinsic magnetic topological materials as demonstrated in MnBi$_2$Te$_4$~\cite{gongyan2019experimental,lijiaheng2019intrinsic}. The intrinsic QAH materials do not require an atomic-scale control of magnetic dopants/alloys or heterojunctions, which are thus preferable for experiments and applications. So far the intrinsic QAH effect has only been observed in the van der Waals (vdW) layered materials MnBi$_2$Te$_4$~\cite{deng2020} and twisted bilayer graphene~\cite{serlin2020}, though interesting theoretical predictions have been made on transition-metal trihalides~\cite{you2019two-2,sun2020intrinsic} and transition-metal dinitrides~\cite{kong2018quantum}. Nevertheless, the FM exchange is rather weak in the existing intrinsic QAH materials, which is the bottleneck for improving working temperature.

The layered iron-based superconductor (FeSC) materials, such as Fe$X$ ($X$=S, Se, Te) and their derivatives, are of particular interest, as rich topological and magnetic properties have been revealed in this material family, e.g., the coexistence of superconductivity with topology~\cite{xu2016topological,zhang2018observation,wang2018evidence}, antiferromagnetism~\cite{lu2015coexistence} or ferromagnetism~\cite{pachmayr2015coexistence}. Remarkably, a superconductivity-ferromagnetism phase transition was observed by injecting massive lithium ions into the layered materials using a state-of-the-art gating technique~\cite{ma2019electric,lei2019phase}. Motivated by the research progresses, we try to utilize lithium injection as a new degree of freedom to tune material properties and seek QAH insulators in FeSC materials. In this work, we use first-principles calculations to show that Li-decorated monolayer Fe$X$ ($X$= S, Se, Te) is a potential candidate for room-temperature ferromagnetic semiconductor as well as high-temperature QAH insulator. In contrast to previous works, we propose a new pathway to create QAH insulators from existing vdW layered materials by using the technique of Li decoration.

The bulk Fe$X$ is a vdW layered material formed by $X$-Fe-$X$ trilayers stacked along the $c$ axis and monolayer Fe$X$, which is the focus of this work, has a tetragonal lattice in the $P4/nmm$ space group. The group IA element Li with an ultralow electronegativity easily loses one valence electron and becomes Li$^+$, making it effective in tuning carrier concentrations. We will mainly discuss the case of FeSe as the decoration of Li has similar influence on the material family.

\begin{figure}
	\includegraphics[width=\linewidth]{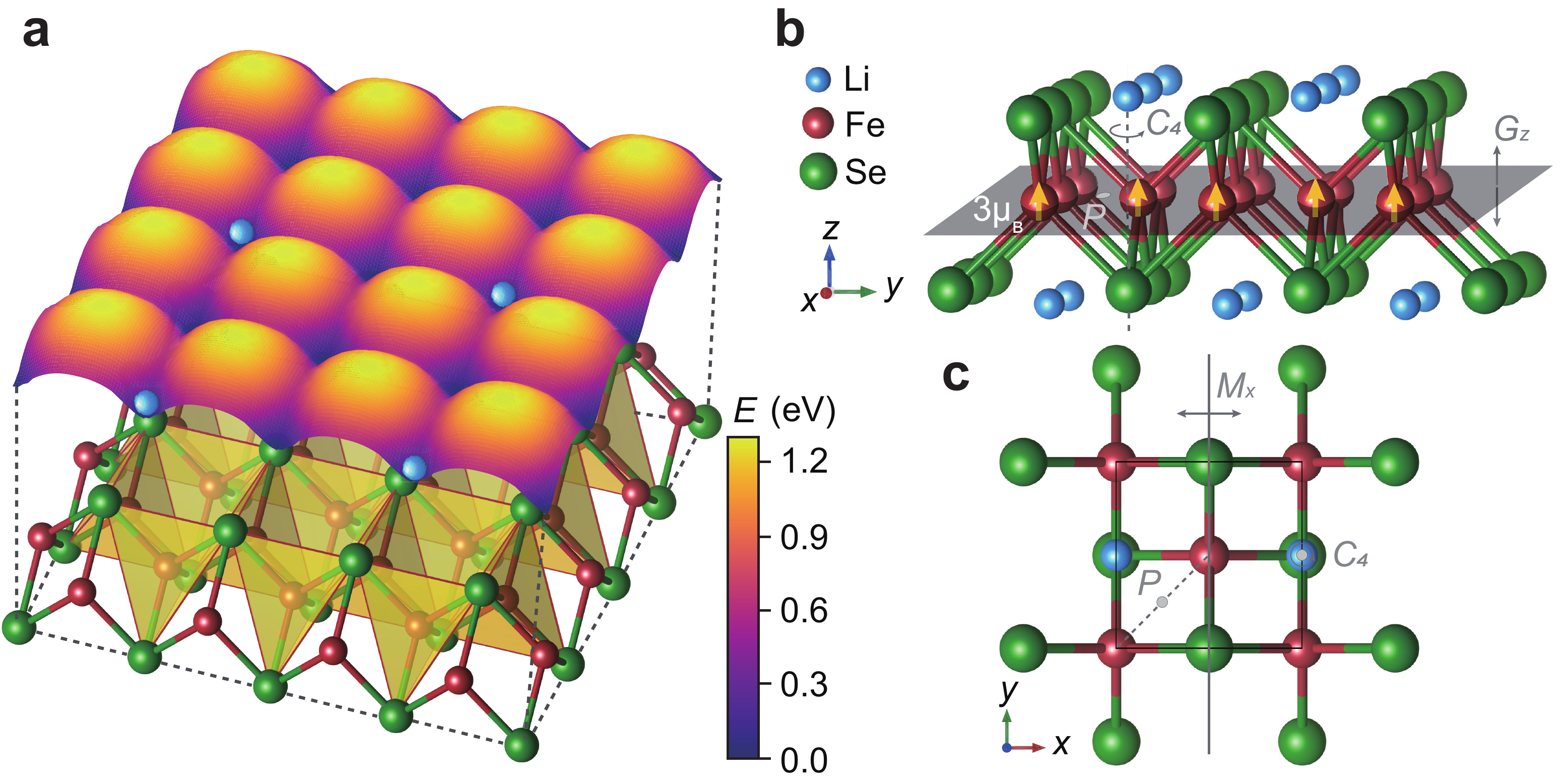}
	\caption{Atomic structure of monolayer LiFeSe. (\textbf{a}) Potential energy surface of Li atoms on monolayer FeSe. (\textbf{b,c}) Side and top views of monolayer LiFeSe. The symmetry operations are displayed. The 2D material has a ferromagneitc ground state with spin magnetic moment of 3.0$\mu_B$ per Fe atom (labelled by the orange arrows). }
	\label{fig1}
\end{figure}

Figure 1A displays the potential energy surface of a Li atom on monolayer FeSe. The potential has an ``egg-box'' like shape, displaying minimums at the hollow sites directly above Se atoms. The hollow sites have a potential energy significantly lower than elsewhere, thus serving as the Li adsorption sites. The more Li atoms added, the more hollow sites occupied. When the ``egg-box'' is fully occupied by Li adatoms, further adding or removing Li atoms would become energetically unfavorable. Thus, a stable material system is expected to emerge in the full occupation limit. This was verified by {\it{ab initio}} random structure searching~\cite{pickard2011ab} (see Methods), which considered 3000 structures and found that almost all of them converged to the same atomic configuration as expected (Fig. 1B,C). This 2D structure is composed of a Li-Se-Fe-Se-Li quintuple layer and shares the same space group with monolayer FeSe. Furthermore, our phonon calculations (Fig. S1) and first-principles molecular dynamic simulations (Fig. S2) show that the structure is dynamically stable and thermally stable above room temperature. We thus conclude that a stoichiometric 2D material LiFeSe is formed by Li decoration of monolayer FeSe.

\begin{figure}
	\includegraphics[width=\linewidth]{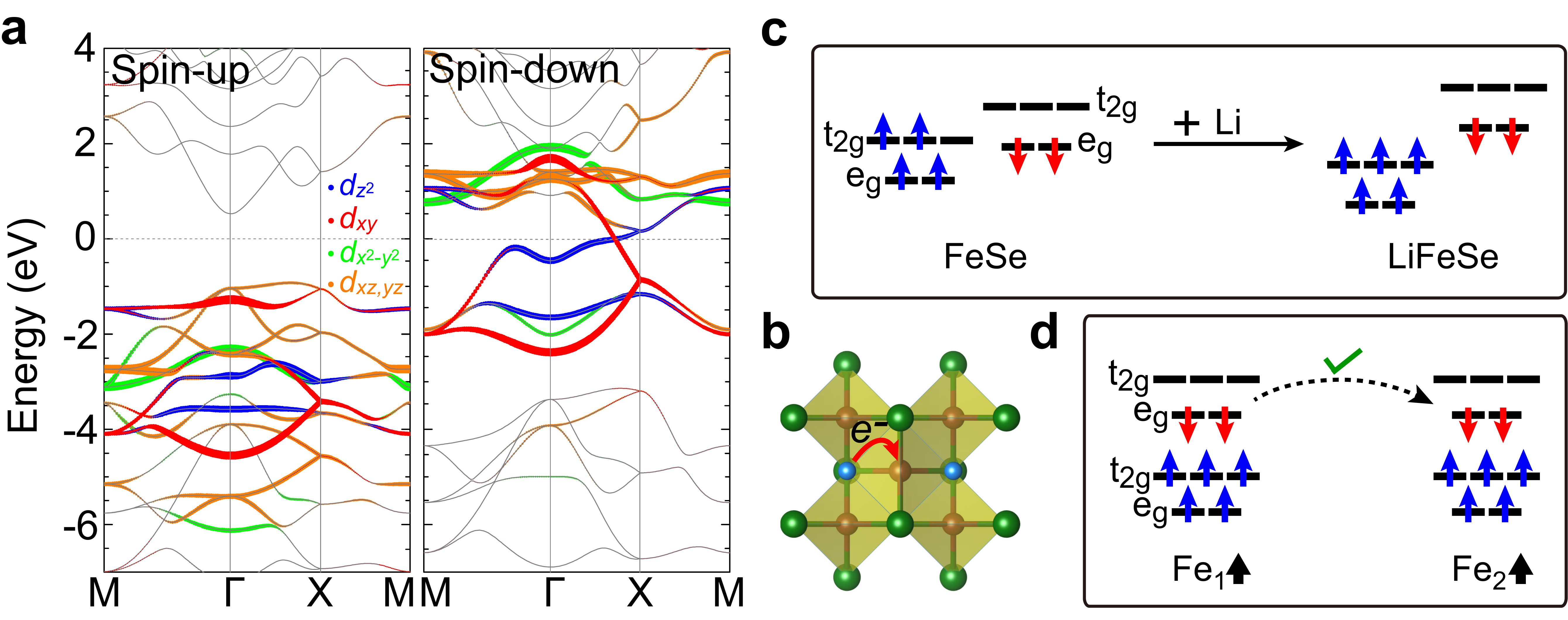}
	\caption{Electronic and magnetic properties of monolayer LiFeSe. (\textbf{a}) Band structure of monolayer LiFeSe without the SOC for spin up (left) and spin down (right) states. The contribution of $d_{z^2}$, $d_{xy}$, $d_{x^2-y^2}$, and $d_{xz,yz}$ orbitals to the Bloch states are denoted by blue, red, green, and orange dots, respectively. (\textbf{b}) Schematic diagram displaying the electron injection from Li to Fe. (\textbf{c}) Schematic diagram of 3$d$ orbital occupations for the Fe atoms in monolayer FeSe and LiFeSe. (\textbf{d}) Schematic diagram of the ferromagnetic kinetic exchange coupling between Fe atoms in monolayer LiFeSe.}
\label{fig2}
\end{figure}

The magnetic properties of monolayer LiFeSe are significantly changed by Li decoration. We investigated various magnetic configurations (Fig. S3) and found that the magnetic ground state is FM (Table S4), in contrast to monolayer FeSe which is proposed to have a checkboard antiferromagnetic (AFM) ground state~\cite{wang2016topological}. More importantly, the FM state is much more stable than other magnetic states. For instance, the calculated energy of FM state is about 0.7 eV per Fe atom lower than that of the checkboard AFM state. The energy difference is found to be on the same order for monolayer LiFeS (0.8 eV/Fe) and LiFeTe (0.6 eV/Fe). The magnetic energy difference is huge, one or two orders of magnitude larger than that of the well known 2D FM materials~\cite{gong2019two}, implying the existence of extremely strong FM exchange coupling here. The magnetocrystalline anisotropy energy (MAE), defined as the total energy difference between in-plane and out-of-plane spin configurations, is calculated to be 0.3, 0.2, -0.7 meV/Fe for LiFeS, LiFeSe and LiFeTe, respectively, while a positive MAE implies an out-of-plane easy axis. Figure 2A displays the electronic structure of monolayer LiFeSe excluding the SOC. The system gives a novel spin polarized state which is a large-gap insulator for spin up and a gapless Dirac semimetal for spin down. This is the so-called 2D half Dirac semimetal state~\cite{pardo2009half,cai2015single,huang2015emergence,you2019two}, which is the parent state of QAH insulators.

To reveal the underlying mechanism of ferromagnetism, we first analysed the orbital occupation of Fe atoms. In LiFeSe, four Se atoms around each Fe atom form a distorted edge-sharing tetrahedron (Fig. 2B), of which crystal field results in an $e_g$-$t_{2g}$ orbital splitting of the $d$ orbitals of Fe, the same as in FeSe. The $d$ orbital occupation of Fe atoms, however, is determined by a competition between the crystal field splitting and the Hund's rule interaction. For FeSe, the low-spin $e_g^4 t_{2g}^2$ state and the high-spin $e_g^3 t_{2g}^3$ state are predicted stable by density functional theory (DFT) and DFT plus dynamical mean-field theory (DFT+DMFT) calculations, respectively~\cite{yin2011kinetic}, indicating that the two types of orbital occupations are possibly close in energy. In contrast, the situation considerably changes in LiFeSe. Because one more electron is injected by each Li (Fig. 2B), the Fe atoms are in the high-spin $e_g^4 t_{2g}^3$ state, giving spin magnetic moment of 3 $\mu_B$ (Fig. 2C). The system thus gets strongly spin polarized.

The strong Fe-Fe interaction is crucial for the ultra-stable FM state. Because of the short distance between Fe atoms (2.6 \AA), the Fe-Fe hybridization is strong even for 3$d$ orbitals, which is also demonstrated through the dispersive 3$d$ states (on the order of eV) near the Fermi level (Fig. 2A). For the Fe atoms in LiFeSe, the spin up channel is fully occupied and the partially occupied spin down channel is at the Fermi level. As a result, the direct electron hopping is only allowed when the magnetic moment of neighboring Fe atoms are parallelly aligned by the Pauli exclusion principle (Fig. 2D). Therefore, the particular orbital occupation of Fe 3$d$ orbitals and the short Fe-Fe distance lead to an ultra-stable FM state in LiFeSe through the strong FM kinetic exchange~\cite{coey2010magnetism,eremin1981kinetic}.

In nature, the Fe bulk has a high $T_C$ of 1043 K. Free-standing monolayer Fe has recently been experimentally fabricated~\cite{zhao2014free}, which could possibly inherit the superior FM properties of the bulk~\cite{vaz2008}. Noticeably, the monolayer Fe and monolayer LiFeSe share similar material traits: The Fe atoms are both in the high-spin states with similar spin magnetic moments (3.1 $\mu_B$ for Fe and 3.0 $\mu_B$ for LiFeSe), and their Fe layers both crystallize in a square lattice with essentially the same Fe-Fe distance ($\sim$2.6 \AA). As discussed above, these features are essential to realize strong FM exchange coupling. Therefore, monolayer LiFeSe is expected to display strong ferromagnetism, in analogy to monolayer Fe~\cite{zhao2014free}. Superior to free-standing monolayer Fe that is not stable on its own, the 2D Fe layer embedded in LiFeSe can stably exist and is not easily affected by environment. By using many-body Green's function method as well as parallel tempering Monte Carlo simulations with {\it{ab initio}} exchange coupling parameters, the Curie temperature of monolayer LiFeSe is estimated to be about 1500 K (Fig. S12), much higher than the existing 2D FM materials (Table S8). As to be shown below, a band gap will be opened by the SOC in the material. This renders monolayer LiFeSe a candidate of the long-sought room-temperature FM semiconductor~\cite{ando2006seeking}.

Symmetry plays an important role in topological physics. The key space-group symmetry operations include space inversion $P$, $C_4$ rotation, $M_x$ and $M_y$ mirrors, and glide mirror $G_z = \{M_z|\frac{1}{2},\frac{1}{2},0\}$ (Fig. 1B,C). The nonsymmorphic symmetry $G_z$ ensures that every band is (at least) doubly degenerate along the high-symmetry X-M and Y-M lines in the Brillouin zone and gets split into separated bands elsewhere. This generally introduces inverted band structures. Specifically, let us focus on the two spin-down bands near the Fermi level, mainly contributed by $d_{z^2}$ and $d_{xy}$ orbitals of Fe, respectively (Fig. 2A). Their band order is inverted between $\Gamma$ and X (or Y), leading to band crossings between valence and conduction bands. The band crossings are protected by $M_y$ ($M_x$), which forbids $d_{z^2}$ and $d_{xy}$ to hybridize. The crossing points, called Dirac points (DPs), appear at generic $\bf{k}$ points along the mirror symmetry invariant lines $\Gamma$-X and $\Gamma$-Y. Actually, there exist two pairs of DPs located at $(\pm k_D,0)$ and $(0, \pm k_D)$, which are named DP$_{\pm 1}$ and DP$_{\pm 2}$ for simplicity. They are related to each other by $P$ and $C_4$.

\begin{figure}
	\includegraphics[width=\linewidth]{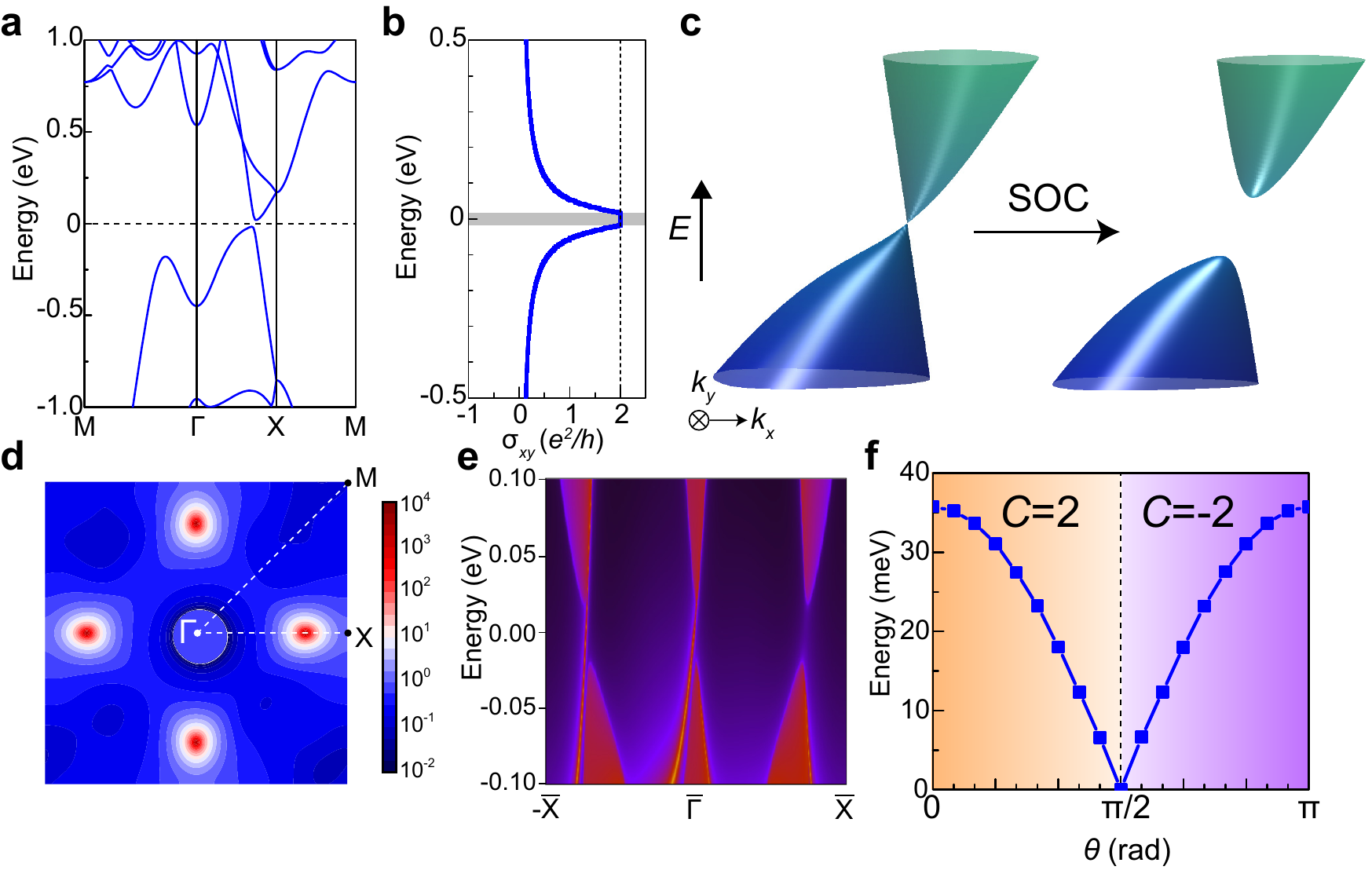}
	\caption{Topological properties of monolayer LiFeSe. (\textbf{a}) Band structure with the SOC. (\textbf{b}) Anomalous hall conductance $\sigma_{xy}$ as a function of Fermi energy, which displays a quantized plateau at the value of 2$e^2/h$ within the bulk gap. (\textbf{c}) 3D band structures of Dirac fermions near the Fermi level excluding (left) and including (right) the SOC. (\textbf{d}) Distribution of Berry curvature contributed by occupied valence bands in the momentum space. (\textbf{e}) Topological edge states calculated along the (100) direction. (\textbf{f}) Dependence of band gap and Chern number on the spin orientation quantified by a polar angle $\theta$, where $\theta = 0, \pi/2$, and $\pi$ denote +$z$, +$x$ and -$z$ directions. }
	\label{fig3}
\end{figure}

To describe the low energy physics of Dirac fermions, we derived a two-band effective Hamiltonian by the $k\cdot p$ method: $H_{\textrm{eff}} = h_0 (\bf{q}) + \bf{h}(\bf{q})\cdot \bm{\sigma}$, where $\bf{q}$ is referenced to the DP, and the Pauli matrices $\bm{\sigma}$ act on the orbital space (see Methods). For DP$_{+ 1}$, $h_0 ({\bf{q}}) = C q_x$, ${\bf{h}(\bf{q})}= (Aq_x, Bq_y, 0)$. An extra term $h_0$ exists due to the low symmetry of the DP. $H_{\textrm{eff}}$ of other DPs (sometimes called valleys) can be obtained by symmetry. In contrast to typical Dirac cones in graphene and TIs, the four Dirac cones in the present system are tilted, anisotropic, and spin polarized (for spin down only) (Fig. 3C), which offers a unique platform to investigate the topological valley physics and the QAH physics.

When including the SOC, $M_x$ and $M_y$ are spontaneously broken by the out-of-plane ferromagnetism. The SOC introduces a Dirac mass term $H' = M \sigma_z$ into the effective Hamiltonian, thus opening a Dirac gap (Fig. 3A,C). We computed the anomalous Hall conductance $\sigma_{xy}$ as a function of Fermi energy (Fig. 3B) and the distribution of Berry curvature in the momentum space (Fig. 3D). For each gapped Dirac cone, the valence states contribute a quantized Berry phase of $\pi$. Multiplying the valley degeneracy of four, a total Berry phase of $4\pi$ is obtained, which corresponds to a quantized anomalous Hall conductance $\sigma_{xy}= Ce^2/h$ with a high Chern number $C=2$. As a hallmark of QAH insulators, two chiral gapless edge modes appear within the bulk gap, as demonstrated by the edge-state calculations (Fig. 3E). Furthermore, if a magnetic domain wall between opposite magnetizations exists in the material, four chiral conduction channels would be topologically protected to appear along the domain boundary. The high-Chern-number states could be used to create exotic elementary excitations~\cite{barkeshli2012topological,wang2013quantum}.

One prominent feature of this material family is that the QAH gap is sizable (35 meV in LiFeSe), larger than the thermal energy at room temperature. The appearance of large QAH gap is attributed to the facts that the effective SOC strength of $d$ orbitals is enlarged by bonding with heavy elements, and that the SOC effects are significantly enhanced near band crossing points. Benefiting from the strong coupling between magnetic and topological states, the QAH gap is highly tunable by magnetic field. By varying the spin orientation from $+z$ to $+x$ then to $-z$ directions, the band gap monotonically decreases down to zero and then reopens (Fig. 3F) as previously studied~\cite{you2019two}. This is accompanied with a topological quantum phase transition from $C=2$ to $C=-2$. Moreover, for generic in-plane magnetizations that break mirror symmetries like $M_{x,y}$, the QAH phase with varying Chern number could be possibly obtained~\cite{liu2013plane}. Such kind of physical processes can be experimentally realized by applying magnetic fields of a few Tesla.

In addition to LiFeSe, we also studied monolayer LiFeS and LiFeTe, which are stable candidate materials for high-temperature QAH insulators as well (Figs. S6 and S11). Note that monolayer LiFeTe is predicted to have in-plane magnetization, which could natively be a 2D half Dirac semimetal. If the magnetization is modulated along the out-of-plane direction (e.g., by exchange bias, strain, magnetic field, Se/Te alloy, etc.), a QAH insulator with a very large gap (70 meV) could be obtained. Furthermore, we investigated the materials MFeSe (M=H, Na, K) and found similar 2D structures only for NaFeSe and KFeSe. These two materials have electronic and magnetic properties similar as LiFeSe, but become metallic due to the existence of nearly free electron bands at the Fermi level (Fig. S8). The structure of H-decorated FeSe is somewhat disordered, possibly due to the small atomic size of H. Therefore, Li is the optimized choice for decorating FeSC materials.

In practice, there exists layered materials composed of Fe$X$ and vdW spacer layers, such as LiOH-FeSe~\cite{lu2015coexistence,woodruff2016parent}. Such kind of vdW materials have broad space for Li diffusion between Fe$X$ layers, advantageous for experimentally fabricating 3D structures of LiFe$X$. Figure 4A displays the atomic structure of 3D LiOH-LiFeSe. Due to the weak vdW coupling between the LiOH and LiFeSe layers, this 3D structure can be viewed as isolated, parallel layers of LiFeSe. This is confirmed by the calculated electronic structure (Fig. 4B), which displays flat bands along the $\Gamma$-Z direction (except for nearly free electron bands). Therefore, the system is formed by stacking 2D QAH insulators together, generating a 3D QAH insulator~\cite{jinyj2018three} with a large gap of 33 meV. The novel topological material can not only display strong Berry-phase effects, but also provide macroscopic number of chiral conduction channels, useful for low-power electronics.

\begin{figure}
	\includegraphics[width=\linewidth]{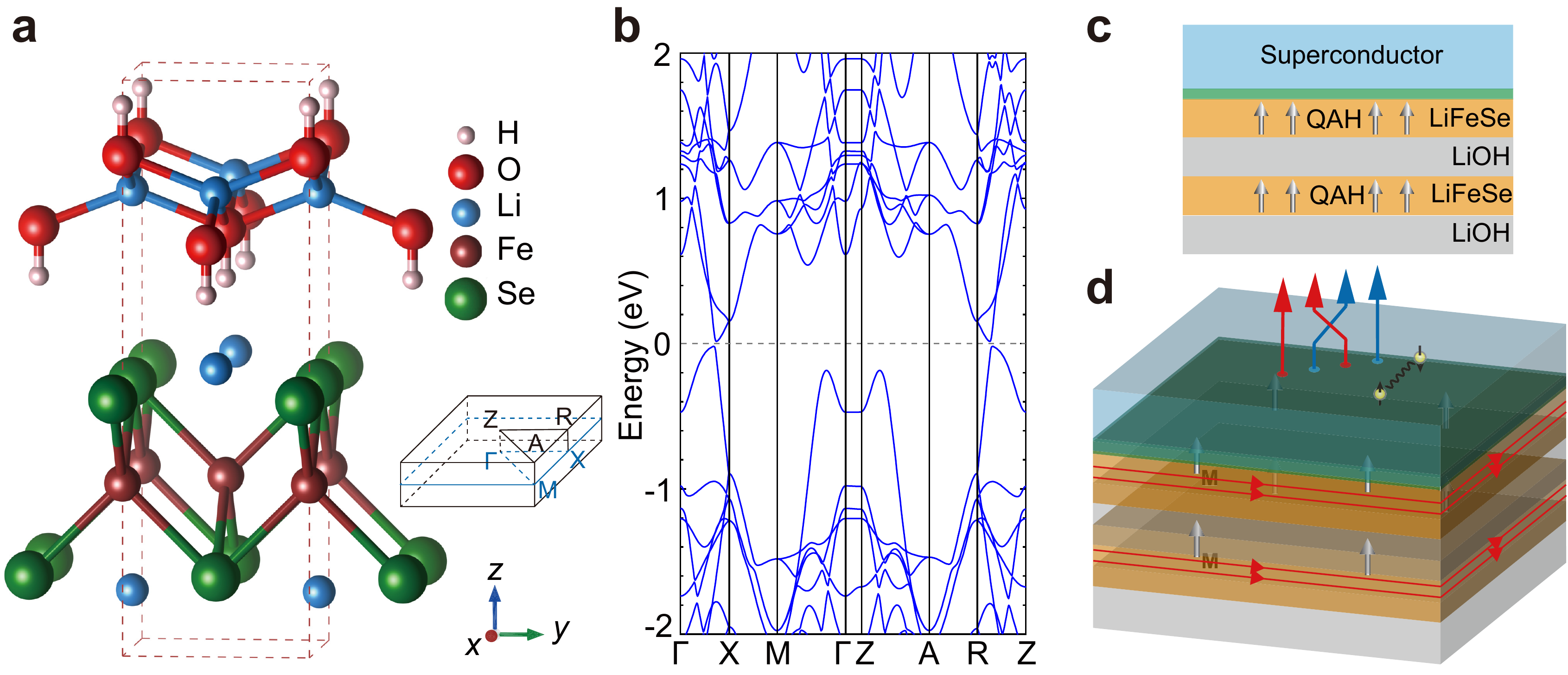}
	\caption{Atomic and electronic structures of 3D LiOH-LiFeSe and potential applications. (\textbf{a}) Atomic structure and the first Brillouin zone of 3D LiOH-LiFeSe. (\textbf{b}) Band structure of 3D LiOH-LiFeSe with the SOC. (\textbf{c,d}) A natural heterostructure between superconductor and QAH insulator (LiFeSe) built by gate-controlled lithium injection, which could be used to detect the chiral topological superconductivity and Majorana fermions. An isolation layer (colored green) is inserted at the interface for blocking the lithium diffusion into the superconductor, which may use thin flakes of graphene or boron nitride.}
	\label{fig4}
\end{figure}

We studied the electronic structure of LiFeSe by using most available computational methods based on DFT (see Methods). All the computation methods (including DFT+DMFT) consistently predicted essentially the same results, including stable valence state Fe$^+$ with spin moment of 3$\mu_B$, strong ferromagnetic exchange coupling, and very similar electronic band dispersions (Figs. S5 and S6). The results clearly indicate that electron correlation effects play a minor role in LiFeSe and support the reliability of the band structure calculations.

It should be remarked that the existence of half Dirac-like band structure is ensured by symmetry and insensitive to computational methods, lattice strains or weak structural disorders, despite some quantitative differences in the predicted magnetization energies and band gaps \cite{supp} (Tables S2-S7 and Figs. S4-S7, S9, and S10). Very recently, experiments have been done on (Li,Fe)OH-Fe$X$ ($X$=S, Se)~\cite{ma2019electric,lei2019phase}. Upon Li injection, a superconducting-ferromagnetic insulating transition and the emergence of ferromagnetism with out-of-plane easy axis and extraordinarily high $T_C$ were observed, but the underlying origin remains mysterious. The key experimental observations can be qualitatively or even quantitatively explained by the first-principles calculation of LiOH-LiFe$X$. Detailed discussions on relevant experiments are presented in the Supplemental Material \cite{supp}.

Finding room-temperature FM semiconductors and seeking high-temperature QAH insulators are among the most challenging problems nowadays, which is of crucial importance to fundamental research and to future development of electronics, spintronics, quantum computation, etc. The strategy to realize high-temperature FM semiconductor and QAH insulator by Li injection is not limited to FeSC materials as we demonstrated here, but generally applicable to other materials, such as nickel-based superconductors. Therefore, one can study the interplay of superconductivity, magnetism and topology in the same material system to explore emergent topological quantum physics. For instance, a natural heterostructure between superconductor and QAH insulator can be fabricated by gate-controlled Li injection, where the chiral topological superconductivity and Majorana fermions could be detected~\cite{qi2010chiral,lian2018topological} (Fig. 4C,D). More critically, many novel quantum effects and devices based on the QAH physics have been theoretically proposed but rarely experimentally observed, because the QAH effect was accessible only at liquid helium temperatures. In this sense, the finding of high-temperature QAH materials opens new opportunities to realize novel practical quantum applications.

\begin{acknowledgments}
This work is supported by the Basic Science Center Project of NSFC (Grant No. 51788104), the Ministry of Science and Technology of China (Grants No. 2016YFA0301001, No. 2018YFA0307100, and No. 2018YFA0305603), the National Natural Science Foundation of China (Grants No. 11674188 and No. 11874035), Tsinghua University Initiative Scientific Research Program, and the Beijing Advanced Innovation Center for Future Chip (ICFC). M.Y. is supported by Shuimu Tsinghua Scholar Program and Postdoctoral International Exchange Program. We acknowledge Changsong Xu for giving the technical support on Monte Carlo simulations and the computational resource provided by Tencent Cloud. 

Y. L. and J. L. contributed equally to this work.
\end{acknowledgments}

\end{document}